\documentclass[preprint2]{aastex}
\usepackage{url}

\usepackage[misc,geometry]{ifsym} 

\usepackage{amsmath}

\RequirePackage{color}

\begin{document}
\title{Accretion-ejection in rotating black holes: a model for `outliers' track of radio-X-ray correlation in X-ray binaries}
\slugcomment{Not to appear in Nonlearned J., 45.}
\shorttitle{Accretion-ejection in rotating black holes}
\shortauthors{Aktar et al.}

\newcommand{\emaila}{sbdas@iitg.ac.in}
\renewcommand\email[1]{\texttt{#1}} 

\author{Ramiz Aktar}
\affil{Indian Institute of Technology Guwahati, Guwahati, 781039, India}

\author{Anuj Nandi}
\affil{Space Astronomy Group, ISITE Campus, U. R. Rao Satellite Center, Outer Ring Road,
	Marathahalli, Bangalore, 560037, India}
\and

\author{Santabrata Das\altaffilmark{1}}
\affil{Indian Institute of Technology Guwahati, Guwahati, 781039, India.}

\altaffiltext{1}{Corresponding author. Email: \email{\emaila}}

\vskip 0.5cm

\begin{abstract}
We study the global accretion-ejection solutions around a rotating black hole considering 
three widely accepted pseudo-Kerr potentials that satisfactorily mimic the space-time geometry of rotating black holes. We find that all the pseudo potentials provide standing shock solutions for large range of flow parameters. We identify the effective region of the shock parameter space spanned by energy ($\mathcal{E}_{\text{in}}$) and angular momentum ($\lambda_{\text{in}}$) measured at the inner critical point ($x_{\text{in}}$) and find that the possibility of shock formation becomes feeble when the viscosity parameter ($\alpha$) is increased. In addition, we find that shock parameter space also depends on the adiabatic index ($\gamma$) of the flow and the shock formation continues to take place for a wide range of $\gamma$ as $1.5 \le \gamma \le 4/3$. For all the pseudo potentials, we calculate the critical viscosity parameter ($\alpha_{\text{shock}}^{\text{cri}}$) beyond which standing shock ceases to exist and compare them as function of black hole spin ($a_k$). We observe that all the pseudo potentials under consideration are qualitatively similar as far as the standing shocks are concerned, however, they differ both qualitatively and quantitatively from each other for rapidly rotating black holes.  
Further, we compute the mass loss from the disc using all three pseudo potentials and find that the maximum mass outflow rate ($R^{\rm max}_{{\dot m}}$) weakly depends on the black hole spin. To validate our model, we calculate the maximum jet kinetic power using the accretion-ejection formalism and compare it with the radio jet power of low-hard state of the black hole X-ray binaries (hereafter XRBs). The outcome of our results indicate that XRBs along the `outliers' track might be rapidly rotating.

\end{abstract}


\section{Introduction}

The accretion of matter around black holes is considered to be the key physical mechanism in understanding the black hole systems. More than four decades ago, \citet{Shakura-Sunayev73} first introduced a standard Keplerian disc model based on self-consistent solutions that successfully explains the thermal component of the X-ray spectrum emitted from the accretion disc around the black hole candidates. But it fails to demonstrate the origin of hard power law tail commonly seen in the observed X-ray spectrum. To address this issue, \citet{Sunyaev-Titarchuk85} proposed a accretion disc model containing Compton cloud which inverse Comptonize the Keplerian soft photons to produce hard X-ray power law tail of the spectrum.
The disc-corona model was extensively studied by numerous group of researchers 
\citep{Burn-Kuperus88, Haardt-Maraschi91, Svensson-Zdziarski94, Tanaka-Lewin95, Poutanen-Svensson96, Zdziarski-etal98, Poutanen-etal17} considering Keplerian flows around the black holes. Meanwhile, \citet{Chakrabarti-Titarchuk95} and \citet{Chakrabarti-Mandal06} showed that the black hole spectral properties are better understood provided the disc is composed of both Keplerian and sub-Keplerian matters. Indeed, in modeling the accretion flow, inner boundary conditions of the black hole demand that the angular momentum of the flow close to the horizon needs to be necessarily sub-Keplerian \citep[and references therein]{Chakrabarti89}. Numerical study also supports this view as the accretion flow enters in to the black hole supersonically \citep{Chakrabarti-Molteni95, Lanzafame-etal98, Giri-etal10, Giri-Chakrabarti13, Sukova-Janiuk2015,Kim-etal17}. Moreover, the above assertions are also endorsed observationally for several black hole candidates as well \citep{Smith-etal01, Smith-etal02, Wu-etal02, Yu-etal04, Smith-etal07, Camiber-Smith13, Debnath-etal2014, Iyer-etal15, Nandi-etal18}.

In an accretion process, rotating inflowing matter starts accreting from the outer edge of the disc with negligible radial velocity. Because of strong gravitational pull of black hole, flow gains it radial velocity as it moves inward and eventually crosses the critical point to become supersonic. Depending on the angular momentum, the flow may have multiple critical points and in that scenario, after crossing the outer critical point, the inflowing matter experiences centrifugal repulsion that causes a virtual barrier in the vicinity of the black hole which triggers the discontinuous transition of flow variables in the subsonic region in the form of shock waves \citep{Chakrabarti89}. Since, black hole does not have any hard boundary, post-shock flow acts as a effective boundary layer around the black hole which is commonly called as post-shock corona (PSC) \citep{Aktar-etal15}. Note that the existence of shock wave in an accretion flow and their astrophysical implications have been extensively studied in the literature both analytically and numerically \citep{Fukue87, Chakrabarti89, Lu-etal99, Becker-Kazanas01, Das-etal01, Fukumura-Tsuruta04, Chakrabarti-Das04, Mondal-Chakrabarti06, Chattopadhyay-Das07, Das-Chattopadhyay08, Kumar-Chattopadhyay13, Das-etal14, Sukova-Janiuk2015,Le-etal16, Aktar-etal17, Sarkar-etal18, Dihingia-etal18}.

Complete understanding of accretion properties around the black holes using full general relativistic calculation is rigorous and complex. The exercise becomes even more difficult in the case of dissipative flow. Fortunately, there exists an alternative approach in terms of pseudo-potential that allows us to utilize the Newtonian concept while retaining the salient features of the black hole space-time geometry. It was \citet{Paczyński-Witta80} who first introduced pseudo-Newtonian potential for Schwarzchild black hole and this potential receives tremendous success in both analytical as well as numerical studies \citep{Chakrabarti89, Narayan-yi94, Molteni-etal94, Molteni-etal96, Machida-etal00, Becker-Kazanas01, Proga-Begelman03, Chakrabarti-Das04, Yuan-etal12a, Yuan-etal12b, Okuda14, Das-etal14, Okuda-Das15, Lee-etal16}. Following the same spirit, several attempts were made to formulate pseudo-Kerr potential for rotating black holes as well \citep{Kerr-1963}. Initially, \citet{Chakrabarti-Khanna92} proposed a pseudo-Kerr potential which is able to replicate the Kerr-geometry at the equatorial plane with reasonable accuracy. Later, \citet{Artemova-etal96} (hereafter ABN96) introduced a prescription for free-fall acceleration around the Kerr black hole. The derivation of pseudo-Kerr potential from this free-fall acceleration is simple and this potential reproduces the features of the Kerr geometry quite well. After that, \citet{Mukhopadhyay02} (hereafter MU02) formulated another pseudo-Kerr potential which is derived in the realm of Kerr space-time geometry. Latter on, \citet{Chakrabarti-Mondal06} (hereafter CM06) prescribed the modified version of the \citet{Chakrabarti-Khanna92} potential which satisfactorily mimics the space time geometry around the rotating black holes of spin $a_k \leq 0.8$. All these pseudo-potentials are formulated and prescribed individually and they have their won limitations to approximate the Kerr space-time geometry. Since the ultimate motivation of these potentials is to describe the space-time geometry around the rotating black hole appropriately, it is essential as well as timely to carry out a comparative study involving all of them. In this context, we consider three different pseudo-Kerr potentials, namely ABN96 \citep{Artemova-etal96}, MU02 \citep{Mukhopadhyay02} and CM06 \citep{Chakrabarti-Mondal06} and study the global transonic accretion flow solutions that contain standing shocks. We compare the shock parameter space spanned by the energy ($\mathcal{E}_{\text{in}}$) and angular momentum ($\lambda_{\text{in}}$) measured at the inner critical point ($x_{\text{in}}$) for inviscid as well as viscous flow. We also compare the critical viscosity parameter calculated using different pseudo potentials that admits standing shock ($\alpha_{\text{shock}}^{\text{cri}}$) and realize that all the pseudo-potentials behave similarly for weakly rotating black holes although they differ considerably when spin of the black hole is increased. Finally, we allow mass loss from the disc and obtain the accretion-ejection solutions. With this, we estimate the maximum outflow rates $(R_{\dot{\text{m}}}^{\text{max}})$ in terms of spin $(a_k)$ of the black hole 
employing the accretion-ejection formalism \citep{Aktar-etal15} for all the pseudo-Kerr potentials.
Thereafter, 
we estimate the maximum kinetic jet power and compare it with the radio-X-ray correlation in black hole X-ray binaries (XRBs) \citep{Corbel-etal13}. Based on this comparative study, we indicate that the black hole XRBs along the `outliers' track are mostly rapidly rotating.

We organize this work as follows. In {\S}2, we present the description of the three pseudo-Kerr potentials. In {\S}3, we describe the assumptions and governing equations for our model. In {\S}4, we discuss the solution methodology and present the results in detail. In {\S}5, we employ our model formalism to estimate the kinetic jet power. Finally, we draw the concluding remarks in {\S}6.

\section{Description of Pseudo-Kerr potentials for black holes}

In this paper, we adopt three different pseudo-Kerr potentials while studying the properties of shock waves around rotating black holes and subsequently compare the obtained results. In the next, we present the detail description of these pseudo-Kerr potentials which are given below.\\ \\
(I) In order to study the properties of accretion flow around rotating black hole, \citet{Artemova-etal96} pro- posed the expression of pseudo-Kerr force which is given by,
$$
F_1(x) = \frac{1}{x^{2-\beta}(x - x_{\text{H}})^{\beta}},
\eqno(1)
$$ 
where $x_{\text{H}}$ is the position of the event horizon and $x$ denotes the radial coordinate. The exact expression of the event horizon is determined from the general relativity \citep{Novikov-Frolov89} as $x_{\text{H}} = 1 + \sqrt{(1 - a_{k}^{2})}$  and the exponent $\beta$ is expressed as $\beta = \frac{x_{\text{ISCO}}}{x_{\text{H}}} - 1$, where $x_{\text{ISCO}}$ stands for the position of the innermost stable circular orbit (ISCO). Following \citet{Bardeen-etal72}, we obtain the expression for innermost stable circular orbit as,
$$
x_{\text{ISCO}} = 3 + Z_{2} \mp [(3 -Z_{1}) (3 +Z_{1} +2 Z_{2})]^{1/2} ,
\eqno(2)
$$
where, $Z_{1} = 1 +(1-a_{k}^{2})^{1/3}[(1+a_{k})^{1/3} + (1-a_{k})^{1/3}]$, and $Z_{2}=(3a_{k}^{2} + Z_{1}^{2})^{1/2}$. Here, `$\mp$' sign stands for prograde and retrograde flow. Here, $a_k$ represents the black hole rotation parameter defined as the specific spin angular momentum of the black hole. In order to obtain the pseudo-Kerr potential $\Phi(x)$, we integrate equation (1) analytically by  imposing the condition $\Phi(x) \rightarrow  0$ for $x \rightarrow \infty$ \citep{Fernandez-etal15} and is given by,
\[
\Phi_1(x)= 
\begin{cases}
\frac{1}{(\beta -1)x_{\text{H}}}\left[1 -\left(\frac{x}{x - x_{\text{H}}}\right)^{\beta -1 }\right] ,& \text{if } \beta\neq 1\\
\frac{1}{x_{\text{H}}}ln\left(1 - \frac{x_{\text{H}}}{x}\right),              & \text{if}  ~\beta = 1
\end{cases} \hfill ~~~~~(3)
\]
for $x > x_{\text{H}}$. The above pseudo-Kerr potential matches exactly with PW80 potential for $a_{k} =0$ and $\beta = 2$. In general, this pseudo-Kerr potential shows good agreement with the result obtained from Kerr geometry. However, for highly spinning black hole, the accretion solutions deviate from the general relativistic results within the limit of 10\% - 20\% error. The general form of the effective pseudo potential ($\Phi_{1}^{\text{eff}}$) is given by,
$$
\Phi_{1}^{\text{eff}} = \frac{\lambda^{2}}{2x^{2}} + \Phi_{1}(x),
\eqno(4) 
$$ 
where the first term in the right hand side denotes the centrifugal potential corresponding to the specific angular momentum of the flow ($\lambda$).\\ \\
(II) \citet{Mukhopadhyay02} formulated the expression of gravitational force $F(x)$ corresponding to the pseudo potential around rotating black hole which is given by,
$$
F_{2}(x) = \frac{(x^{2} - 2a_{k} \sqrt{x} + a_{k}^{2})^{2}}{x^{3}\left[\sqrt{x}(x - 2) + a_{k}\right]^{2}}.
\eqno(5)
$$
The above pseudo-Kerr force successfully reproduces the inner disk properties which are in close agreement with the Kerr geometry for moderately spinning black holes. In case of rapidly rotating black holes, accretion solution deviates from the general relativistic results although the error remains restricted within the acceptable limit of 10\%. The corresponding expression of the pseudo potential ($\Phi_2(x)$) is obtained as,
$$
\Phi_{2}(x) = \int_{\infty}^{x} {F_{2}(x) dx}.
\eqno(6)
$$
It is to be noted that $\Phi_{2}(x)$ reduces to the PW80 potential for $a_k = 0$.

Similar to equation (4), we obtain the effective pseudo-Kerr potential as,
$$
\Phi_{2}^{\text{eff}} = \frac{\lambda^{2}}{2x^{2}} + \Phi_{2}(x).
\eqno(7) 
$$ \\ \\
(III) \citet{Chakrabarti-Mondal06} supplemented an alternative pseudo-Kerr effective potential that satisfactorily captures the general relativistic features around black hole for $a_k \lesssim 0.8$. The expression of the effective pseudo-Kerr potential ($\Phi_{3}^{\text{eff}}$) is given by,
$$
\Phi_{3}^{\text{eff}} = -\frac{B+\sqrt{B^2-4AC}}{2A},
\eqno(8)
$$
where,
$$
A=\frac{\epsilon^2 \lambda^2}{2x^2},
$$

$$
B=-1 + \frac{\epsilon^2 \omega \lambda r^2}{x^2} 
+\frac{2a_k\lambda}{r^2 x},
$$

$$
C=1-\frac{1}{r-x_0}+\frac{2a_k\omega}{x}
+\frac{\epsilon^2 \omega^2 r^4}{2x^2}.
$$
Here, $x$ and $r$ represent the cylindrical and spherical radial distance. Here, $x_0=0.04+0.97a_k+0.085a_k^2$, $\omega=2a_k/(x^3+a^2_k x+2a^2_k)$ and $\alpha^2=(x^2-2x+a^2_k)/(x^2+a_k^2+2a^2_k/x)$, $\epsilon$ is the redshift factor. 
The corresponding pseudo-Kerr force is obtained as $F_{3}(x) \equiv$ $\Phi_{r}^{'}=\left(\frac{\partial{\Phi_{3}^{\text{eff}}}} {\partial{r}}\right)_{z<<x}$, where, $z$ is the vertical height in the cylindrical coordinate system and $r=\sqrt{x^{2}+z^{2}}$. In the next section, we present the governing equations that describe the inflowing and outflowing matter around a rotating black hole.

\section{Modeling of Accretion Disc}

We consider a steady, advective, viscous, axisymmetric accretion flow around a rotating black hole. Here, we consider the disc is confined around the equatorial plane and the jet or outflow geometry is considered in the off-equatorial plane about the axis of rotation of the black hole \citep{Molteni-etal96, Chattopadhyay-Das07,Aktar-etal17}. For simplicity, we adopt pseudo-Kerr approach to describe the space-time geometry around rotating black holes. In order to express the flow variables, we consider an unit system as $G = M_{\rm BH} = c =1$ throughout the paper. In this unit system, radial coordinate, angular momentum and velocity are computed in units of $GM_{\rm BH} /c^2$, $GM_{\rm BH}/c$, and $c$, respectively.

\subsection{Governing Equations for Accretion}

Here, we present the hydrodynamical equations that govern the accretion flow around the rotating black holes and are given by,\\
(i) The radial momentum conservation equation:
$$
v\frac{dv}{dx} + \frac{1}{\rho}\frac{dP}{dx} + \frac{d\Phi_{i}^{\text{eff}}}{dx} = 0,
\eqno(9)
$$
where $v$, $P$, $\rho$ and $x$ represent the radial velocity, isotropic gas pressure, density and radial distance of the flow, respectively. Here, $\Phi_{i}^{\text{eff}}$ is the effective pseudo-Kerr potential around black hole and the subscript $i$ can take any one value among 1, 2, and 3 depending on the choice of the pseudo-potentials. We define the adiabatic sound speed as $a = \sqrt{\gamma P/\rho}$, where $\gamma$ represents the adiabatic index. In this work, we use $\gamma = 1.4$ all throughout unless otherwise stated.

(ii) The mass conservation equation:
$$
\dot{M}=4 \pi \rho v x h(x) ,
\eqno(10)
$$
where $\dot{M}$ denotes the mass accretion rate which is a global constant throughout the flow except the region of mass loss and $4\pi$ is the geometric constant. Here, $h(x)$ refers to the half-thickness of the flow. Considering the hydrostatic equilibrium in the vertical direction for thin disc, we calculate the half-thickness of the disc as,
$$
h(x)=a\sqrt{\frac{x}{\gamma F_{i}(x)}},
\eqno(11)
$$
where $F_{i}(x)$ represents the pseudo-Kerr force corresponding to the pseudo-Kerr potential described in $\S$2.\\
(iii) The angular momentum distribution equation:
$$
v\frac{d\lambda}{dx} + \frac{1}{\Sigma x}\frac{d}{dx}(x^2 W_{x\phi}) =0 ,
\eqno(12)
$$
where $W_{x\phi}$ is the $x\phi$ component of the viscous stress tensor. Following \citet{Chakrabarti96b}, we consider the expression of $W_{x\phi}$ as,
$$
W_{x\phi}^{(1)} = -\alpha (W + \Sigma v^2).
\eqno(13)
$$
where $\alpha$ denotes the viscosity parameter. Here, $W~(= 2I_{n+1} P h)$ and $\Sigma~(= 2I_n \rho h)$ represent the vertically integrated pressure and density. Here, $I_{n}$ and $I_{n+1}$ are the constant factors of integration of vertically averaged density and pressure \citep{Matsumoto-etal84} where $I_n = (2^n n!)^2 /(2n + 1)!$ and $n~[= 1/(\gamma - 1)]$ is the polytropic index.  \\
Finally,\\
(iv) The entropy generation equation:
$$
\Sigma v T \frac{ds}{dx} = Q^{+} - Q^{-},
\eqno(14)
$$
where T is the temperature and s is the entropy density of the accretion flow, respectively. In addition, $Q^{+}$ and $Q^{-}$ represent the heat gain and heat lost by the flow. In this work, for the purpose of simplicity, we ignore cooling effect and consequently we choose $Q^{-} = 0$. After some simple algebra, equation (14) becomes,
$$
\frac{v}{\gamma -1}\left[\frac{1}{\rho}\frac{dP}{dx} - \frac{\gamma P}
{\rho^2}\frac{d\rho}{dx}\right]=-\frac{Q^{+}}{\rho h}= -H.
\eqno(15)
$$
Using the mixed shear stress prescription \citep{Chakrabarti96b, Aktar-etal17}, we calculate the heating of the flow by means of viscous dissipation as,
$$
Q^{+} = A{\rho h} (g a^2 + \gamma v^2) \left(x\frac{d\Omega}{dx}\right),
\eqno(16)
$$
where, $A = -\frac{2\alpha I_n}{\gamma}$ and $g = \frac{I_{n+1}}{I_n}$. 

\subsection{Critical Point Conditions}

In the process of accretion on to black hole, inflowing matter starts its journey subsonically from the outer edge of the disk and eventually enters into the black hole with supersonic speed. This scenario evidently demands that accretion flow must change its sonic state from subsonic to supersonic at some point between the outer edge of the disc and the black hole horizon. Such a special point is called as critical point where accretion flow maintains certain conditions. In order to calculate these critical point conditions, we make use of equations ($9-16$) to obtain the velocity gradient which is given by,
$$
\frac{dv}{dx} = \frac{N}{D},
\eqno(17)
$$
where,
\begin{align*}
N =& -\frac{A\alpha  (ga^2 +  \gamma v^2)^2}{\gamma  vx} - \frac{3 a^2 v}{(\gamma -1)x} \nonumber \\ & +\frac{ a^2 v}{(\gamma 
	-1)}\left(\frac{d\ln F_{i}(x)}{dx}\right) - \frac{3 A \alpha g a^2 
	(ga^2 + \gamma v^2)}{\gamma vx}   \\ &+ \left[\frac{2A\alpha g  (ga^2 + \gamma v^2)}{v} + 
\frac{(\gamma +1)v}{(\gamma -1)}\right]\left(\frac{d\Phi_{i}^{\text{eff}}}
{dx}\right)  
\nonumber \\& + \frac{A \alpha g a^2(ga^2 +  \gamma v^2)}
{\gamma v}\left(\frac{d\ln F_{i}(x)}{dx}\right) 
\\& + \frac{2 A\lambda (ga^2 + \gamma v^2)}
{x^2} , \tag{17a}
\end{align*}
and 
\begin{align*}
D = & \frac{2 a^2}{(\gamma -1)} - \frac{(\gamma + 1)v^2}{(\gamma - 1)} \\ &
- A \alpha (ga^2 + \gamma v^2) \left[(2 g - 1) - \frac{g a^2}{\gamma v^2}\right]. \tag{17b}
\end{align*}

Using equation (12) and (17), we calculate the gradient of angular momentum as,
\begin{align*}
\frac{d\lambda}{dx} =& \frac{\alpha}{\gamma v}(ga^2 +  \gamma v^2) + 
\frac{2\alpha x ga}{\gamma v}\left(\frac{da}{dx}\right) \\
&+ \alpha x \left(1
- \frac{ga^2}{\gamma v^2}\right)\left(\frac{dv}{dx}\right). \tag{18}
\end{align*}
Further, we calculate the gradient of sound speed using equations ($9-11$) as,
\begin{align*}
\frac{da}{dx} =& \left(\frac{a}{v}-\frac{\gamma v}{a}\right)\frac{dv}{dx} + 
\frac{3a}{2x} - \frac{a}{2}\left(\frac{d\ln F_{i}(x)}{dx}\right)  \\
&- 
\frac{\gamma}{a}\left(\frac{d\Phi_{i}^{\text{eff}}}{dx}\right). \tag{19}
\end{align*}
As discussed, the accreting matter around black hole is smooth everywhere along the flow streamline and therefore, the radial velocity gradient must be real and finite always. However, depending on the flow variables, $D$ may vanish at some radial coordinate. Since $dv/dx$ remains smooth always, the point where $D$ tends to zero, $N$ must also vanish there. Such a point where both $N$ and $D$ simultaneously goes to zero is identified as critical point and $N = D = 0$ are the critical point conditions. Setting $D = 0$, we find the radial velocity of the flow ($v_c$) at the critical point ($x_c$) as,
$$
v^2_c = \frac{-m_b - \sqrt{m_b^2 - 4 m_a m_c}}{2 m_a}a^2_c,
\eqno(20)
$$
where $a_c$ is the sound speed at $x_c$ and
\begin{align*}
m_a =&-A\alpha  \gamma^2 (\gamma - 1)(2g - 1) - \gamma (\gamma +1), \\ 
m_b =& 2\gamma - 2A \alpha g \gamma (\gamma -1)(g - 1),\\ 
m_c =&A\alpha g^2(\gamma -1).
\end{align*}

	Setting $N=0$, we get an algebraic equation of sound speed ($a_c$) as,
	$$
	a_1 a^2_c + a_2 a_c + a_3 =0,
	\eqno(21)
	$$
where
\begin{align*}
a_1 =& -\frac{A\alpha (g +  \gamma M_c^2)^2}{\gamma x_c} - \frac{3 M_c^2}{(\gamma -1)x_c} 
\\ &+ \frac{M_c^2}{(\gamma -1)}\left(\frac{d\ln F_{i}(x)}{dx}\right)_c - \frac{3A \alpha g (g +  \gamma M_c^2)}{\gamma x_c}  \\ & + \frac{A \alpha g(g + \gamma M_c^2)}
{\gamma}\left(\frac{d\ln F_{i}(x)}{dx}\right)_c,\\
a_2 =&  \frac{2A\lambda M_c (g + \gamma M_c^2)}
{x^2_c},{\rm and} \\
a_3 =& \left[2A\alpha g (g + \gamma M_c^2) + \frac{(\gamma + 1)M_c^2}{(\gamma - 
	1)}\right]\left(\frac{d\Phi_{i}^{\text{eff}}}{dx}\right)_c.                
\end{align*}
Here, $M_c$ refers the Mach number at $x_c$, where Mach number of the flow is defined as $M = v/a$.
We solve equation (21) to calculate $a_c$ and consider only the positive root of the equation (21) as $a_c>0$ always. The detail steps to obtain $a_c$ from equation (21) is given in appendix-A.

The nature of the critical point is determined by the value of $dv/dx$ at $x_c$ \citep[and reference therein]{Das07}. At the critical point, $dv/dx = 0/0$ and therefore, we apply l'Hospital rule to calculate $(dv/dx)_c$. Usually, $(dv/dx)_c$ possesses two values. When both the derivatives are real and of opposite sign, the critical point is called as saddle type critical point and any physically acceptable accretion solution can only pass through it. When shock forms, accretion flow passes through two saddle type critical points: one in the pre-shock region and the other in the post-shock region \citep[and reference therein]{Chakrabarti-Das04}. In the subsequent sections, we refer the saddle type critical point as critical point only. In general, critical points in the post-shock flow form very close to the horizon and called as inner critical points ($x_{\text{in}}$). On the other hand, critical points in the pre-shock flow usually form far away from the black hole and called as outer critical points ($x_{\text{out}}$).

\subsection{Standing Shock Conditions}

In order to form standing shock, accreting flow variables must satisfy the Rankine-Hugonoit (RH) shock conditions \citep{Landau-Lifshitz59} which are given by,

\noindent (i) the conservation of energy flux:

The specific energy of the flow ($E$) is given by \citep{Becker-etal08, Das-etal09},
$$
E = \frac{v^2}{2} + \frac{a^2}{\gamma - 1} - \frac{\lambda^2}{x^2} + \frac{\lambda \lambda_{\rm H}}{x^2} + \Phi_{i}^{\rm eff},
$$
where $\lambda_{\rm H}$ denotes the angular momentum of the flow at the event horizon. Since energy conservation is preserved across the shock front, using $E_{+}=E_{-}$, we obtain
$$
\mathcal{E}_{+} = \mathcal{E}_{-},
\eqno(22a)
$$
where the subscripts `$-$' and `$+$' indicate the flow variables just before and after the shock, respectively. Here, $\mathcal{E}(x)$ denotes the local specific energy of the flow equivalent to the canonical Bernoulli parameter and is calculated as $\mathcal{E}(x) = v^2/2 + a^2 /(\gamma -1) + \Phi^{\text{eff}}_i$. It may be noted that while obtaining equation (22a), we use $\lambda_{+}=\lambda_{-}$ across the shock front.

(ii) the conservation of mass flux:
$$
\dot{M}_{+} = \dot{M}_{-} - \dot{M}_{\text{out}} = \dot{M}_{-}(1 - R_{\dot{\text{m}}}),
\eqno(22b)
$$
where $\dot{M}_{+}$ and $\dot{M}_{-}$ represent the accretion rates across the shock front, respectively. The outflow rate is defined as $R_{\dot{\text{m}}} = \dot{M}_{\text{out}}/\dot{M}_{-}$. \\
Finally,\\
(iii) the conservation of momentum flux:
$$
W_{+} + \Sigma_{+}v_{+}^2 = W_{-} + \Sigma_{-}v_{-}^2,
\eqno(22c) 
$$
where, $W$ and $\Sigma$ are the vertically integrated pressure and density as described earlier \citep[and references therein]{Das-etal01}. 

\subsection{Equations for Outflow and Computation of Mass Loss}

Due to the shock transition, the post-shock flow becomes very hot and dense and eventually, PSC acts as an effective boundary around the black hole. As a result, a part of the accreting matter is deflected by PSC and driven out in the vertical direction by the excess thermal gradient force across the shock, producing bipolar outflows (\citet{Chakrabarti99, Chattopadhyay-Das07, Das-Chattopadhyay08}, and reference therein). To calculate the mass outflow rates, we employ the formalism 
adopted by \citet{Aktar-etal15}. As the jets are tenuous in nature, we ignore viscosity in the outflowing matter. We also consider that the outflowing matter obey the polytropic equation of states, i.e., $P_{j} = K_{j}\rho_{j}^{\gamma}$, where subscript `$j$' 
refers the jet variables and 
$K_{j}$ represents the measure of specific entropy of the jet, respectively. The equations of motion for the outflow are given below. 

\noindent (i) The energy conservation equation of outflow:
$$
\mathcal{E}_{j} = \frac{v_j^2}{2} + \frac{a_{j}^2}{\gamma -1} + \Phi_{i}^{\text{eff}},
\eqno(23)
$$
where $\mathcal{E}_{j}$, $v_{j}$ and $a_{j}$ are the specific energy, velocity and sound speed for the outflowing matter, respectively. $\Phi_{i}^{\text{eff}}$ is the effective pseudo-Kerr potentials mentioned in section \S2.

\noindent (ii) The mass conservation equation of outflow:
$$
\dot{M}_{\text{out}} = \rho_{j}v_{j}\mathcal{A}_{j},
\eqno(24)
$$
where $\dot{M}_{\text{out}}$ and $\mathcal{A}_{j}$ are the outflowing rate of mass and area function for the jet, respectively. We calculate $\mathcal{A}_{j}$ by knowing the radius of two boundary surfaces, namely centrifugal barrier (CB) and funnel wall (FW) \citep{Molteni-etal96}. The radius of CB is obtained using pressure maximum surface i.e., $(d\Phi_{i}^{\text{eff}}/dx)_{r_{\text{CB}}} = 0$ and the radius of FW is defined as the pressure minimum surface, i.e., $\Phi^{\rm eff}_{i}\vert_{r_{\rm FW}}=0$ \citep{Molteni-etal96, Aktar-etal15, Aktar-etal17}.  We also consider the projection factor $\sqrt{1+(dx_{j}/dy_{j})^2}$ for calculating jet area function  \citep{Kumar-Chattopadhyay13, Aktar-etal17}. 

As the outflow is originated from the PSC region, we assume that the outflow is essentially launched with the same density as in the PSC, $i.e.$, $\rho_{j} = \rho_{+}$. Therefore, using the equations (10), (22b) and (24), we calculate the mass loss rate as,
$$
R_{\dot{\text{m}}} = \frac{Rv_j\mathcal{A}_{j} \sqrt{\gamma F_{i}}}
{4\pi a_{+}v_{-}x^{3/2}_{s}},
\eqno(25)
$$  
where $R$ is the compression ratio defined as $R=\Sigma_{+}/\Sigma_{-}$. Further, $v_{j}$, $\mathcal{A}_{j}$ and $F_{i}$ denote the jet velocity, jet area function and pseudo-Kerr force calculated at the shock $x_s$, respectively. We use the successive iterative method to calculate $R_{\dot{\text{m}}}$ as described in \citet{Aktar-etal15}.

\section{Results}

\subsection{Global Accretion Solutions including Shock}

	In order to obtain the global accretion solution around the black holes, the inner boundary conditions demand that at the horizon, the flow radial velocity approaches the speed of light and the viscous stress vanishes. Keeping these in mind, we choose a set of flow variables, namely, critical point ($x_c$), angular momentum at $x_c$ $(\lambda_{c})$ and viscosity parameter ($\alpha$), and simultaneously integrate equations ($17-19$) from the critical point in the outward direction. When the flow reaches to a large distance representing the outer edge of the disc ($x_{\text{edge}}$), we again integrate equations ($17-19$) from the critical point up to close to the horizon. Finally, we join these two parts of the solution to get a complete global transonic accretion solution around the black holes, provided the radial velocity of the flow becomes comparable to the speed of light just outside the horizon. Here, we avoid to check the vanishing of the viscous stress at the horizon, simply because the adopted pseudo potential approach is generally poorly valid near the event horizon.
Further, we note the values of all the flow variables at $x_{\text{edge}}$. In actuality, we would get the identical accretion solution obtained above, when equations ($17-19$) are solved using the flow variables at the outer edge of the disc.

In Section 3.2, we point out that shocked accretion flow must contains two critical points. In reality, during the course of accretion, subsonic accretion flow from the outer edge of the disc first crosses the outer critical point ($x_{\text{out}}$) to become supersonic and continues to accrete towards the black hole. Meanwhile, centrifugal repulsion becomes dominant in the vicinity of the black hole and hence, inflowing matter is forced to be slowed down there. Effectively, a virtual centrifugal barrier is formed that triggers the discontinuous transitions of flow variables in the subsonic region which is commonly known as shock transition. For standing shock transition, RH shock conditions need to be satisfied (see \S3.3). After the shock transition, flow gradually attains its speed due to the strong gravitational pull and ultimately enters into the black hole supersonically after passing through the inner critical point ($x_{\text{in}}$). In this subsection, we consider no mass loss from the disk i.e., $R_{\dot{\text{m}}} = 0$.

\begin{figure}[t]
	\begin{center}
		\includegraphics[angle=00,width=0.48\textwidth]{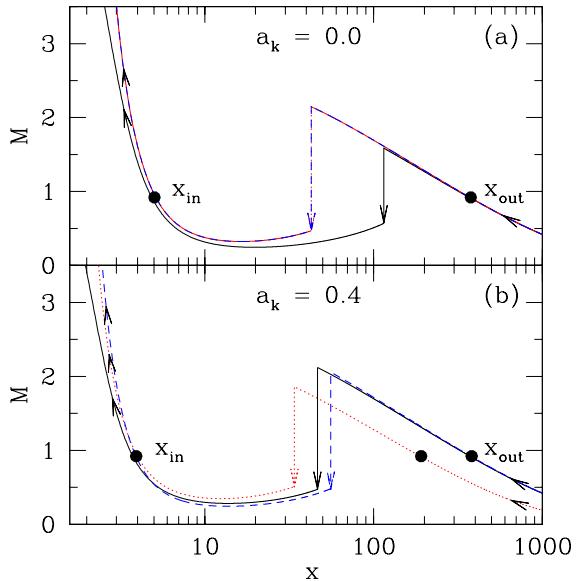}
	\end{center}
	\caption{Illustration of shocked accretion solution where the variation of Mach number ($M=v/a$) is shown with radial distance ($x$). In the upper panel, results are shown for non-rotating ($a_k = 0$) black hole whereas in the lower panel, $a_k = 0.4$ is chosen. Solid, dotted and dashed curves represent the solutions obtained for CM06, MU02 and ABN96 potentials, respectively. Here, we fix $\gamma = 1.4$. See text for details.}
\end{figure}

In Fig. 1, we compare the shock induced global accretion solutions obtained using different pseudo-Kerr potentials. Here, the input parameters of the flow are kept fixed at the outer edge of the disc. In the upper panel (Fig. 1a), we choose the outer edge of the disc as $x_{\text{edge}} = 1000$ and inviscid accreting flow is injected from $x_{\text{edge}}$ with energy $\mathcal{E}_{\text{edge}}$ = 0.001 and $\lambda_{\text{edge}}$ = 3.35 on to a non-rotating black hole. Solid, dotted and dashed curves represent the results obtained for CM06, MU02 and ABN96 potentials where the vertical arrows indicate the location of shock transitions at 115.07 for CM06 and at 42.84 for both MU02 and ABN96 potentials. In the case of non-rotating black hole ($a_k$), since MU02 and ABN96 potential become identical, accretion solutions for these two potentials display complete overlap all throughout. In the lower panel (Fig. 1b), we choose $a_k = 0.4$ and compare the shocked accretion solutions for three different potentials considering the same set of inflow parameter fixed at $x _{\text{edge}}$ except $\lambda_{\text{edge}}$. Here, we fix $x_{\text{edge}} = 1000$, $\mathcal{E}_{\text{edge}} = 0.001$, $\lambda_{\text{edge}} = 2.98$ and $\alpha = 0$. As before, solid, dotted and dashed curves denote the results corresponding to CM06, MU02 and ABN96 potentials and the respective shock locations are calculated as 46.56 (CM06), 33.99 (MU02), and 55.79 (ABN96), respectively. From the figure, it is clear that even for the same set of input parameters, the adopted potentials display noticeably different results as far as the shock transition is concerned. This possibly happens due to the fact that these potentials are primarily approximated and they tentatively mimic the space-time geometry around the rotating black holes. In both panels, inner critical point ($x_{\text{in}}$) and outer critical point ($x_{\text{out}}$) are marked with filled circles and overall direction of the flow motion is indicated by arrows.

\begin{figure}[t]
	\begin{center}
		\includegraphics[angle=00,width=0.48\textwidth]{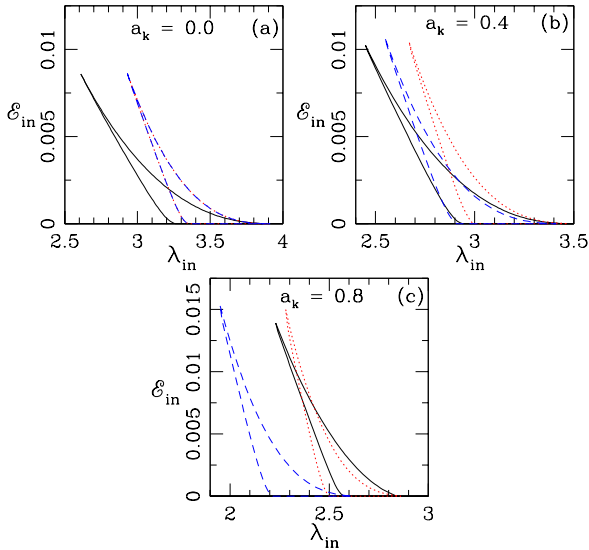}
	\end{center}
	\caption{Classification of shock parameter space for three different pseudo-Kerr potentials. Here,  inviscid flow ($\alpha = 0.0$) is considered for three different spin values ($a_k=0.0, 0.4$ and $0.8$) which are marked in each panel. Solid, dotted and dashed curves represent results for CM06, MU02 and ABN96 pseudo-Kerr potentials, respectively. Here, we fix $\gamma = 1.4$. See text for details.}
\end{figure}

It is generally believed that in the context of understanding the black hole spectral properties \citep{Chakrabarti-Mandal06} as well as jets and outflows \citep{Das-Chakrabarti08, Aktar-etal15, Sarkar-Das16, Aktar-etal17}, shock induced global accretion solutions are potentially preferred over the shock free solutions. Therefore, it is worthy to identify the range of flow parameters that admits shocks. Towards this, in Fig. 2, we compute the shock parameter space spanned by the energy ($\mathcal{E}_{\text{in}}$) and angular momentum ($\lambda_{\text{in}}$) of the inviscid flow measured at the inner critical point ($x_{\text{in}}$). In the figure, we fix the spin values as $a_k$ = 0.0 $(a)$, 0.4 $(b)$ and 0.8 $(c)$, respectively and in each panel, region bounded by the solid, dotted and dashed curves are obtained for CM06, MU02 and ABN96 pseudo-Kerr potentials. As expected, in Fig. 2a, the shock parameter spaces for MU02 and ABN96 potentials are overlapped. This is obvious because MU02 and ABN96 potentials exactly reduce to same potential form for $a_k= 0.0$ as mentioned earlier. But, the shock parameter space for CM06 significantly differs from the same obtained for the remaining two potentials although a common overlapping region is found. In Fig. 2b, we choose $a_k$ = 0.4 and observe that the shock parameter spaces deviate from each other for all the potentials. Interestingly, here also a common region among the parameter spaces is found. These common regions are particularly important to compare the accretion solutions among different adopted potentials (see Fig. 1). Moreover, we observe that the parameter spaces shift towards higher energy and lower angular momentum domain with the increase of the black hole spin ($a_k$) for all the potentials. This apparently indicates that the accretion flow continues to sustain standing shock around rapidly rotating black holes provided its energy is relatively high. When the black hole spin is further increased as $a_k = 0.8$, shock parameter space for ABN96 is significantly shifted to the low angular momentum side and completely separated from the rest leaving any short of common union with others.

\begin{figure}[t]
	\begin{center}
		\includegraphics[angle=00,width=0.48\textwidth]{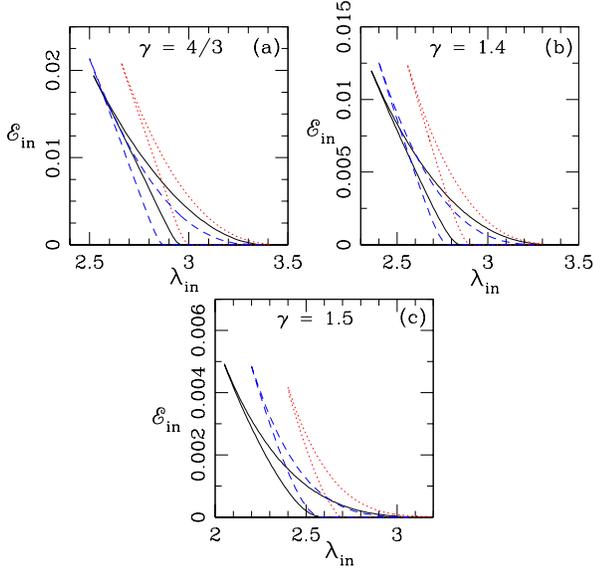}
	\end{center}
	\caption{Comparison of shock parameter space in $\lambda_{in}$ - $\mathcal{E}_{in}$ plane for different $\gamma$ values. Region separated using solid, dotted and dashed curves are obtained for CM06, MU02 and ABN96 pseudo-Kerr potentials, respectively. Here, we consider $\alpha$ = 0 and $a_k$ = 0.5. In each panel, the value of $\gamma$ is marked. See text for details.}
\end{figure}

Until now, we have regarded the accreting matter to be adiabatic in nature and the flow is characterized by an adiabatic index having a representative value $\gamma$ = 1.4. However, in reality, the acceptable theoretical limit of the adiabatic index lies in the range $4/3 \leq \gamma \leq 5/3$ \citep{Frank-etal02}. In order to understand the role of the $\gamma$ values in deciding the global accretion solutions containing standing shock, we study the shock parameter space as function of $\gamma$ for all the potentials. While doing this, the accretion flow is considered to be of three types, namely thermally ultra-relativistic $(\gamma \sim 4/3)$, thermally trans-relativistic ($\gamma \sim 1.4$) and thermally semi-non-relativistic ($\gamma \sim 1.5$), respectively \citep{Kumar-etal13, Aktar-etal15} and obtain the shock parameter space as shown in Fig. 3. Here, we choose, $a_k$ = 0.5 and $\alpha$ = 0 and the obtained results are plotted in Fig. 3 where in each panel, solid, dotted and dashed curves represent the results corresponding to CM06, MU02 and ABN96 potentials. Also, $\gamma$ values are marked in each panel. We find that for a given $\gamma$, the effective region of parameter spaces are different from each other for all the three potentials. In addition, we observe that as the $\gamma$ value is increased, the shock parameter spaces shift towards the lower angular momentum and lower energy sides irrespective to the any chosen form of potential. What is more is that effective region of the parameter space is shrunk as $\gamma$ value is increased. This essentially indicates that the possibility of shock formation is reduced when the flow moves towards non-relativistic limit \citep{Aktar-etal15}.

\begin{figure}[t]
	\begin{center}
		\includegraphics[angle=00,width=0.48\textwidth]{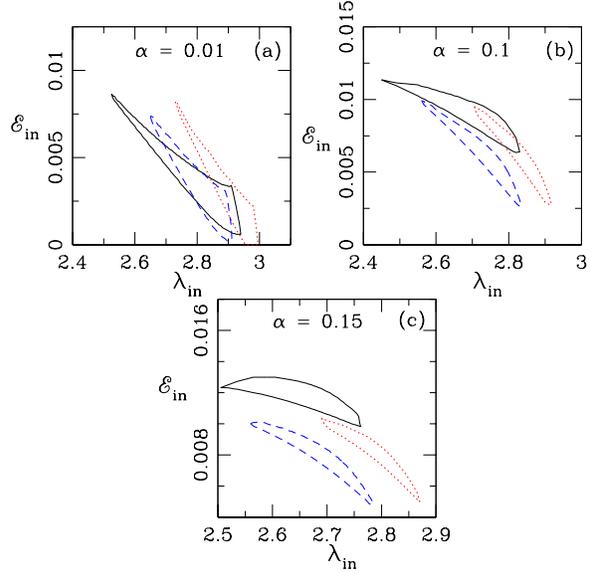}
	\end{center}
	\caption{Modification of shock parameter space for dissipative accretion flow in $\lambda_{\text{in}} - \mathcal{E}_{\text{in}}$ plane. Effective region bounded with solid, dotted and dashed curves are calculated for CM06, MU02 and ABM96 pseudo-potential, respectively. Here, the results are obtained considering $a_k$ = 0.4 and $\gamma = 1.4$. In each panel viscosity parameter is marked. See text for details.}
\end{figure}

\begin{figure}[t]
	\begin{center}
		\includegraphics[angle=00,width=0.48\textwidth]{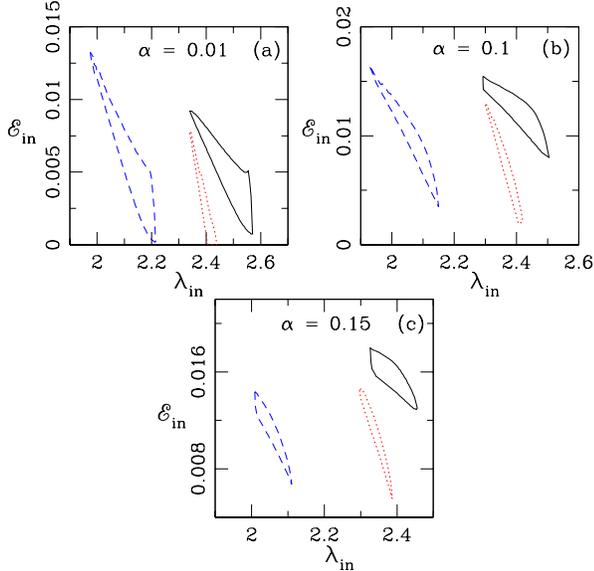}
	\end{center}
	\caption{Same as Fig. 4 but black hole spin is chosen as $a_k$ = 0.8.}
\end{figure}

So far, we have studied the shocked accretion solutions for non-dissipative flow. In our subsequent analysis, we relax this criteria and consider the viscous dissipation process to be active in the flow. With this, we calculate the standing shock parameter space for all the adopted potentials in terms of viscosity parameter ($\alpha$) and display the results in Fig. 4 and Fig. 5. We choose $a_k$ = 0.4 in Fig. 4 and $a_k$ = 0.8 in Fig. 5 and in both figures, vary the viscosity parameter as $\alpha$ = 0.01 $(a)$, 0.1 $(b)$ and 0.15 $(c)$, respectively. In each panel, solid, dotted and dashed curves represent the results corresponding to CM06, MU02 and ABN96 potentials, respectively. Inside the disc, viscosity plays dual role; in one hand viscosity transports angular momentum outward reducing its value at the inner edge and in the other hand, viscous dissipation causes the heating of the flow as it accretes. Because of this, as viscosity is increased, standing shock parameter space is overall shifted towards the higher energy and lower angular momentum side for all the potentials. Moreover, the increase of $\alpha$ introduces enhanced viscous dissipation inside the flow and therefore, the possibility of shock formation is reduced \citep{Chakrabarti-Das04, Das07, Aktar-etal17} which is being realized as the effective region of the parameters space is shrunk when the value of the $\alpha$ parameter is increased. However, it is not possible to increase $\alpha$ indefinitely, because beyond a critical limit ($\alpha_{\text{shock}}^{\text{cri}}$), shock solutions disappears completely.

\begin{figure}[t]
	\includegraphics[angle=00,width=0.48\textwidth]{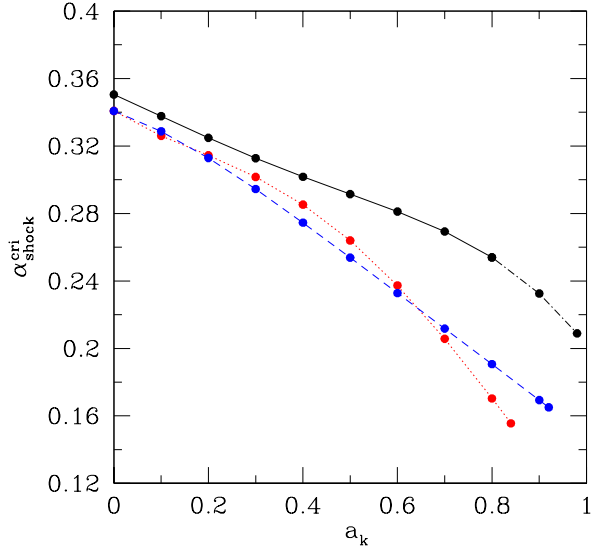}
	\caption{Variation of critical viscosity parameter ($\alpha_{\text{shock}}^{\text{cri}}$) for shock as function of black hole spin ($a_k$). Filled circles joined with solid, dotted and dashed lines represent results obtained using CM06, MU02 and ABN96 pseudo-potentials, respectively. For CM06, we extend the calculation of $\alpha_{\text{shock}}^{\text{cri}}$ beyond $a_k > 0.8$ to examine the overall trend and show the result using dot-dashed curve. Here, we choose $\gamma$ = 1.4. See text for details.}
\end{figure}

Further, we calculate the critical viscosity parameter ($\alpha_{\text{shock}}^{\text{cri}}$) that allows standing shock solutions and plot the variation of $\alpha_{\text{shock}}^{\text{cri}}$ with the spin parameter ($a_k$) for three different potentials, as depicted in Fig. 6. Here, filled circles connected by solid lines, dotted lines and dashed lines are for CM06, MU02 and ABN96 potentials, respectively. While calculating $\alpha_{\text{shock}}^{\text{cri}}$ for a fixed $a_k$, we freely vary the flow parameters, namely $x_{\text{in}}$, $\mathcal{E}_{\text{in}}$ and $\lambda_{\text{in}}$, respectively. Usually, in the weak viscosity limit, the sub-Keplerian flow joins with Keplerian disc quite far away from black hole. Hence, the possibility of finding standing shock which requires the existence of multiple critical points increases at the lower viscosity range. On the contrary, when $\alpha > \alpha_{\text{shock}}^{\text{cri}}$, Keplerian disc approaches very close to the black hole resulting the flow to pass through the inner critical point only \citep{Chakrabarti96b} without having a shock. We find that $\alpha_{\text{shock}}^{\text{cri}}$ is anti-correlated with $a_k$ for all the potentials. Note that we calculate shock solutions for CM06 potential considering rapidly rotating black hole ($a_k \rightarrow 0.98$) as well, although this potential bears limitation to mimic the Kerr geometry satisfactorily for $a_k > 0.8$. Certainly, this introduces error in our calculation, however, it provides us the overall trend of $\alpha_{\text{shock}}^{\text{cri}}$ variation towards the highest value of $a_k$. In case of MU02 and ABN96 potentials, no such restriction is imposed on the upper limit of $a_k$ values. But, we do not find standing shock solutions beyond $a_k > 0.84$ for MU02 and $a_k > 0.92$ for ABN96 potentials, respectively. In addition, we observe that $\alpha_{\text{shock}}^{\text{cri}}$ obtained from different potentials possesses close by values for weakly rotating black holes and it starts deviating from each other with the increase of $a_k$.

\begin{figure}[t]
	\includegraphics[angle=00,width=0.48\textwidth]{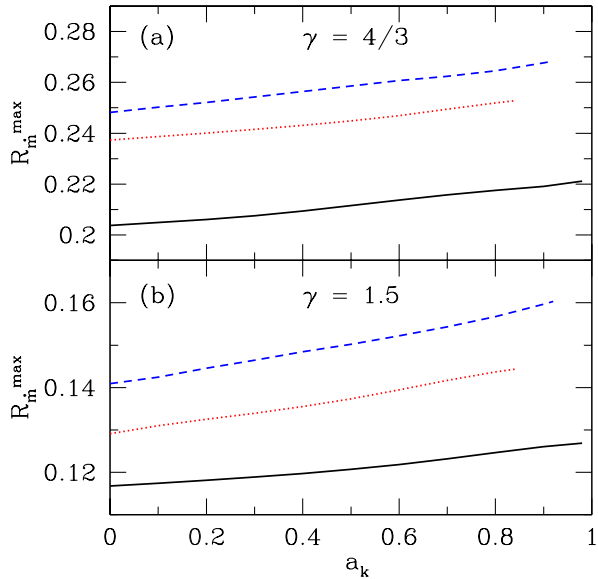}
	\caption{Variation of maximum outflow rates $R_{\dot{m}}^{\text{max}}$ with the black hole spin $a_k$. Upper panel (a): for $\gamma = 4/3$ and lower panel (b): for $\gamma = 1.5$, respectively. Solid, dotted and dashed curves are calculated for CM06, MU02 and ABM96 pseudo-potentials, respectively. Here, viscosity parameter is chosen as $\alpha = 0.05$. See text for details.}
\end{figure}

\subsection{Estimation of Maximum Outflow Rates}

So far, we have performed a comparative study of the accretion flows using pseudo-Kerr potentials where mass loss from the disc is ignored. In reality, due to the shock transition, a part of the inflowing matter is emerged out from the disc as outflow. Rigorous investigations including mass loss from the disc around rotating black hole have already been performed by \citet[and references therein]{Aktar-etal15, Aktar-etal17, Aktar-etal18} using pseudo-Kerr potential \citep{Chakrabarti-Mondal06}. 
In this work, we carry out a comparative study of maximum mass outflow rates $(R_{\dot{\text{m}}}^{\text{max}})$ in terms of black hole spin ($a_k$) using different pseudo-Kerr potentials to examine their effectiveness. 
Employing the accretion-ejection model formalism, we self-consistently calculate the mass outflow rates $(R_{\dot{m}})$ by supplying the inflow parameters, namely flow energy $(\mathcal{E}_{\rm in})$, flow angular momentum $(\lambda_{\rm in})$, viscosity parameters $(\alpha)$, adiabatic index $(\gamma)$ and spin $(a_k)$ of the black hole. 
Now, we freely vary all the inflow parameters and calculate $R_{\dot{\text{m}}}^{\text{max}}$ for a particular $a_k$ \citep{Aktar-etal15, Aktar-etal17}. In Fig. 7, we show the
variation of $R_{\dot{\text{m}}}^{\text{max}}$ with $a_k$ for viscous flow ($\alpha = 0.05$). Here, we choose the two extreme limit of adiabatic index, namely $\gamma =4/3$ that corresponds to thermally ultra-relativistic flow (upper panel) and $\gamma=1.5$ representing the thermally semi-non-relativistic (lower panel) \citep{Aktar-etal15}. For $\gamma = 4/3$,
we find that $R_{\dot{\text{m}}}^{\text{max}}$ corresponding to CM06, MU02 and ABN96 lies in the range $20.37 - 22.11\%$, $23.73 - 25.28\%$ and $24.81 - 26.83\%$, respectively. On the other hand, when $\gamma = 1.5$ is chosen, the value of  $R_{\dot{\text{m}}}^{\text{max}}$ belongs to the range $11.67 - 12.68\%$, $12.91 - 14.43\%$ and $14.09 - 16.03\%$ for CM06, MU02 and ABN96 potentials.

Overall, we realize that the ultra-relativistic ($\gamma = 4/3$) flows produce more outflows compared to the semi-non-relativistic $(\gamma = 1.5)$ flows as far as the maximum outflow rates are concerned. This happens due to the fact that in this work, outflows are purely thermally driven. It may also be noted that ABN96 pseudo potential effectively provides more $R_{\dot{\text{m}}}^{\text{max}}$ compared to other two potentials. Moreover, we observe that $R_{\dot{\text{m}}}^{\text{max}}$ depends on $a_k$ very weakly for all the potentials. 
With this findings, we argue that the correlation between black hole spin and powering jets seems to be feeble. 
It may be noted that the value of $R_{\dot{\text{m}}}^{\text{max}}$ allows us to compute the  kinetic jet power $(L_{\text{jet}}^{\text{est}})$ for black hole sources \citep{Aktar-etal15, Nandi-etal18}. 

In the next section, we apply our accretion-ejection formalism to estimate the kinetic jet power and attempt to explain the observed radio jet power in the low-hard state of the black hole XRBs.

\section{Astrophysical Application}

\begin{figure*}[t]	
	\includegraphics[angle=00,width=0.999\textwidth]{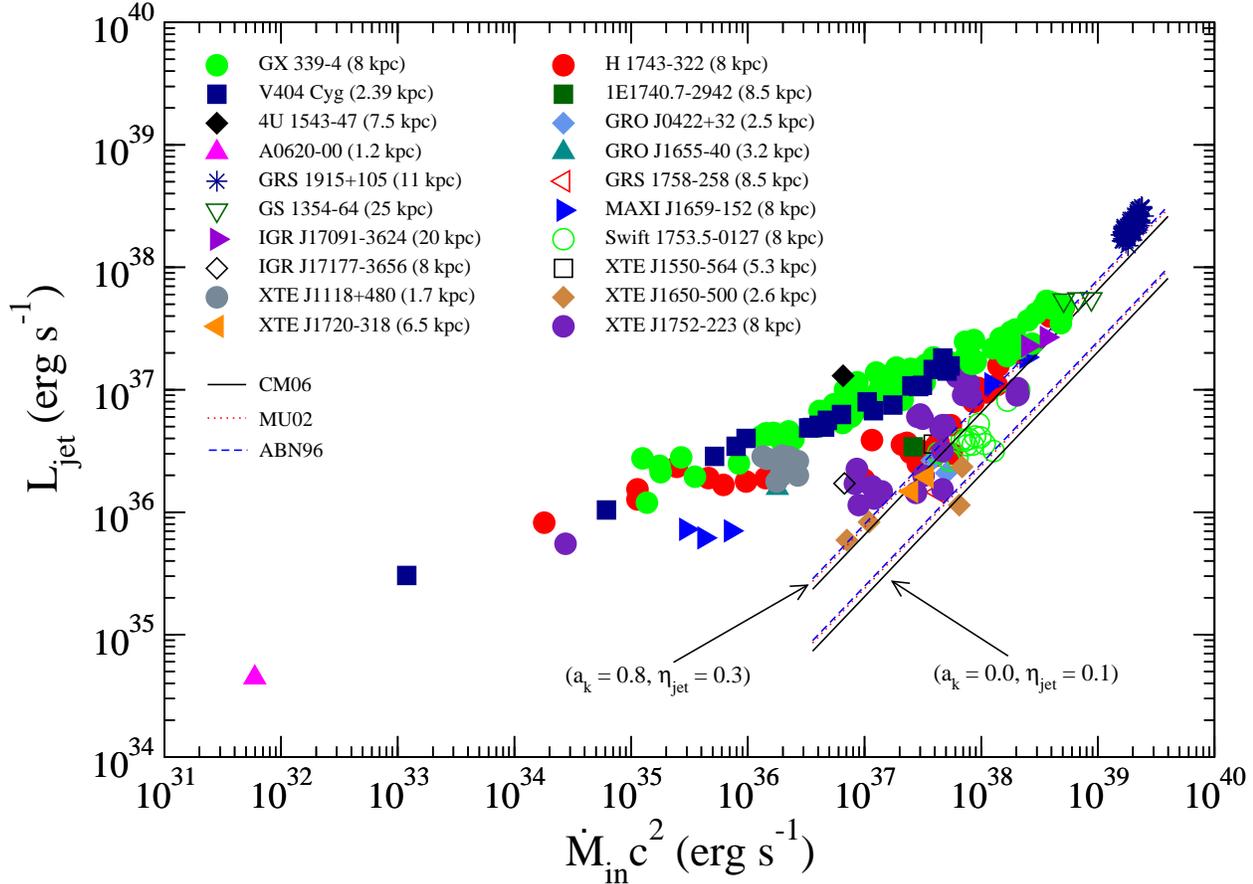}
	\caption{Comparison of observed and theoretical kinetic jet power as a function of accretion power. The different symbols and colors represent the data of low-hard state of 20 black hole XRBs which are taken from \citet{Corbel-etal13}. Length scale mentioned within the parenthesis indicates the distance of the source. The corresponding solid, dotted and dashed lines represent the maximum kinetic jet power from theoretical model for CM06, MU02 and ABN96 potentials, respectively. Chosen values of ($a_k, \eta_{\rm jet}$) used in model calculations are marked. See the text for details.
}
\end{figure*}

\subsection{X-ray and radio correlation of XRBs}

\cite{Fender-etal2005,Fender-etal2009} reported the existence of radio-X-ray correlation in the low-hard states of the XRBs. Interestingly, most of the XRBs follow a universal non-linear correlation, namely $F_R \propto F_X^b$, where $b\sim 0.5 - 0.7$ and $F_R$ and $F_X$ denotes radio and X-ray fluxes, respectively \citep{Hannikainen-etal98, Corbel-etal00, Corbel-etal03, Gallo-etal03, Corbel-etal13}. However, a growing number of sources $e.g.$, H1743-322, Swift 1753.3-0127, XTE J1650-500, XTE J1752-223 are found to lie well outside the universal radio-X-ray correlation \citep{Jonker-etal10, Coriat-etal11, Cadolle-Bel-etal07, Soleri-etal10, Corbel-etal04, Ratti-etal12,Huang-etal14} following an `outliers' track. These sources follow a steeper correlation as $b\sim1.4$ 
\citep{Coriat-etal11}.

\subsection{Kinetic jet power of steady-compact jets: theory and observation}

In this section, we compare the theoretically obtained kinetic jet power with observations. While doing that we 
convert the observed radio luminosity to jet power.
The empirical relation between radio luminosity and jet power is computed considering a simple conical jet model of optically thick jet as \citep{Blandford-Konigl79, Falcke-Biermann96, Heinz-Sunyaev03},
$$
L_R \propto L_{\rm jet}^{17/12},
\eqno(26)
$$
where $L_R~(= 4\pi d^2 F_R)$ is the radio luminosity measured at frequency $\nu$, $F_R$ is the radio flux measured at frequency $\nu$ and $d$ is the distance of the source, respectively. Later, \citet{Heinz-Grimm05} identifies a relation between the jet power and radio luminosity for Cyg X-1 and GRS 1915+105 considering the normalization factor $\sim 6.1\times 10^{-23}$ \citep{Huang-etal14} as,
$$
L_{\rm jet} = 4.79\times 10^{15} L_R^{12/17}~{\rm erg~s^{-1}}.
\eqno(27)
$$

In the low-hard states, the jets are not relativistically boosted and thus we ignore Doppler correction while estimating jet power \citep{Gallo-etal03}. In the present analysis, we employ equation (27) to estimate the kinetic jet power from radio luminosity for all the sources under consideration.
We also calculate the accretion power by using X-ray luminosity $(L_X)$ as $\dot{M}_{\rm in}c^2 = L_X/\eta_{\rm acc}$, where $\eta_{\rm acc}$ is the accretion efficiency factor and $L_X = 4\pi d^2 F_X$, $F_X$ being the X-ray flux.
We obtain $F_X$ ($1-10$ keV) and $L_R$ ($8.6$ GHz) fluxes for the various sources from \citet[references therein]{Corbel-etal13} and plotted in Fig. 8. The different symbols and colors represent the different sources. It is noteworthy that the spin value of some of the selected sources is not yet settled. Hence, for simplicity, we choose $\eta_{\rm acc} = 0.15$ while calculating the accretion power for all the selected sources \citep{Frank-etal02, Longair11}, presented in Fig. 8.

Employing our accretion-ejection model formalism, we compute the maximum kinetic jet power \citep{Aktar-etal15} as,
$$
L_{\rm jet}^{\rm max} = \eta_{\rm jet}\times R_{\dot{m}}^{\rm max} \times \dot{M}_{\rm in} \times c^2~ {\rm erg~s^{-1}},
\eqno(28)
$$
where, $R_{\dot{m}}^{\rm max}$ is the maximum outflow rates and $\eta_{\rm jet}$ is the jet efficiency factor. For the purpose of representation, we choose $\alpha = 0.05$ and $\gamma = 4/3$ and calculate $R_{\dot{m}}^{\rm max}$ for non-rotating ($a_k = 0.0$) and rapidly rotating ($a_k = 0.8$) black holes, respectively (see Fig. 7). In this analysis, we consciously restrict ourselves to choose $a_k \le 0.8$, as one of the adopted potential (CM06) fails to describe space-time geometry satisfactorily above this limiting range of spin value.

We compare our theoretical results (equation 28) with observation (equation 27) which is shown in Fig. 8. 
The solid, dotted and dashed curves represent the theoretically obtained kinetic jet power ($L_{\rm jet}^{\rm max}$) for CM06, MU02 and ABN96 potentials, respectively where the lower curves are for non-rotating black holes ($\eta_{\rm jet} = 0.1$) and the upper curves are for rapidly rotating black holes ($\eta_{\rm jet} = 0.3$), as depicted in Fig. 8. For $a_k = 0.0$, maximum outflow rates are computed as $R_{\dot{m}}^{\rm max} = 0.2037$ (CM06), $0.2373$ (MU02) and $0.2481$ (ABN96) whereas $R_{\dot{m}}^{\rm max} = 0.2175$ (CM06), $0.2519$ (MU02), $0.2645$ (ABN96) for $a_k = 0.8$, respectively. It is clear that $L_{\rm jet}^{\rm max}$ roughly remains insensitive on the choice of potential. And, finally we observe that the `outliers' track \citep[references therein]{Corbel-etal13} agrees quite consistently with the model predictions for rapidly rotating black holes. 

\section{Concluding Remarks}

In this work, we present a comparative study of the accretion-ejection solutions including shock wave by adopting three pseudo potentials prescribed by \citet{Artemova-etal96}, \citet{Mukhopadhyay02} and \citet{Chakrabarti-Mondal06}. These potentials are known to describe the space-time geometry of rotating black holes quite satisfactorily. The advantage of using pseudo-Kerr potentials in lieu of the general theory of relativity (GTR) is that it allows us to investigate the accretion flow properties following the Newtonian approach (i.e. avoiding the rigorous mathematical complexity of GTR) while retaining all the salient features of complex space time geometry around a rotating black hole. Utilizing these potentials, we present the generalized governing equations that describe the dissipative accretion flow around the rotating black hole. We then simultaneously solve these equations to obtain the global transonic accretion solutions and employing the Rankine-Hugoniot shock conditions, we further obtain the shock induced global accretion solutions around a rotating black hole. 

We find that standing shock continues to form in all the adopted pseudo-Kerr potentials (see Fig. 1). We also observe that shocked solutions are not the discrete solutions, instead a wide range of flow parameters admits shock transition in the accretion flow variables. In this context, we identify the effective region of the parameter space spanned by the energy ($\mathcal{E}_{\text{in}}$) and the angular momentum($\lambda_{\text{in}}$) of the flow measured at the inner critical points that allows standing shock solutions and find that shock forms around weakly rotating as well as rapidly rotating black holes (see Fig. 2). Further, we examine the role of adiabatic index ($\gamma$) in determining the shock solutions and notice that the shock parameter space is squeezed when $\gamma$ is tending to the thermally non-relativistic limit (see Fig. 3). This provides a hint that the formation of standing shock is more likely for flows with lower $\gamma$ values.

We continue the study of shock parameter space considering dissipative accretion flow and compare the parameter space in terms of viscosity parameter ($\alpha$). We find that parameter space is gradually modified and shrunk with the increase of $\alpha$ for all the pseudo potentials (see Fig. $4-5$). This evidently indicates that the possibility of shock formation is reduced as the viscous dissipation is enhanced. Beyond a critical limit ($\alpha > \alpha_{\text{shock}}^{\text{cri}}$), accretion flow fails to satisfy the standing shock conditions and therefore, shock disappears completely. Needless to mention that $\alpha_{\text{shock}}^{\text{cri}}$ does not bear any universal value, but depends on the other input parameters (see Fig. 6). In case of weakly rotating black holes, $\alpha_{\text{shock}}^{\text{cri}}$ for all the pseudo-potentials agrees quite well, but differs considerably for rapidly rotating black holes. Hence, we argue that as far as the standing shocks are concerned, qualitatively all the pseudo-potentials behave quite similarly, but they differ both qualitatively and quantitatively from each other for rapidly rotating black holes. Moreover, we realize that CM06 potential provides standing shock solutions for $a_k \rightarrow 0.98$ although this potential ensues erroneous results for $a_k >0.8$ as it fails to describe the space-time geometry beyond this limit. In comparison, we do not find standing shock solutions beyond $a_k > 0.84$ for MU02 and $a_k > 0.92$ for ABN96 potentials (see Fig. 6). 

We further compare the maximum outflow rates $(R_{\dot{\text{m}}}^{\text{max}})$ in terms of the black hole spin ($a_k$) for all the adopted pseudo potentials considering viscous accretion flow ($\alpha = 0.05$). We find that there exist a feeble correlation between $R_{\dot{\text{m}}}^{\text{max}}$ and spin $a_k$ irrespective to the choice of potentials although ABN96 potential provides more $R_{\dot{\text{m}}}^{\text{max}}$ compared to the other potentials (Fig. 7).

We apply our accretion-ejection model to explain the `outliers' track of the X-ray-radio correlations in black hole XRBs. We select sources in their low-hard states from \citet{Corbel-etal13}. We find that theoretical results obtained for the rapidly rotating black holes are in agreement with the observational findings of the black hole XRBs lying along the `outliers' track (see Fig. 8). 

Finally, we point out that the present model bears some limitations. For example, we adopt pseudo potentials to describe the gravitational effect around rotating black hole. Moreover, in our accretion-ejection model, outflows are mainly thermally driven although, in reality, the jet generation from the vicinity of the rotating black holes is likely to be steered by the large scale magnetic fields \citep{Blandford-Znajek77}, radiation pressure \citep{Chattopadhyay-etal2004}, etc. In addition, jet power tends to follow non-linear relation with the accretion power \citep{Huang-etal14, Ghisellini-etal14} as well. All these 
issues may cause our theoretical estimate different from  
the `universal' track except for few sources characterized with higher accretion rate. Although the above issues seem to be relevant in the context of jet generation, we ignore them in the present analysis for the purpose of simplicity. We plan to continue our investigation including them in a future work and will be reported elsewhere.

\section*{Acknowledgments}

Authors are thankful to the anonymous reviewer for useful comments and suggestions that help to improve the manuscript. Authors express sincere gratitude to S. Corbel for sharing the observational data with them. 
AN thanks GD, SAG; DD, PDMSA and Director, URSC for encouragement and continuous support to carry out this research.
 
\section*{Compliance with Ethical Standards}
\begin{itemize}
	\item The authors declare that they have no potential conflicts of interest
	\item This work does not involve Human Participants and/or Animal
\end{itemize}

\appendix 

\section{Calculation of sound speed ($a_c$) at the critical point ($x_c$)}

Putting $N = 0$ in equation (17a) and using equation (20), we get an algebraic equation of $a_c$ which is given by,
$$
 \frac{A\alpha  \left(g+  \gamma \frac{v_c^2}{a_c^2}\right)^2a_c^4}{\gamma  x_c} + \frac{3 a_c^2 v_c^2}{(\gamma -1)x_c} -\frac{ a_c^2 v_c^2}{(\gamma -1)}\left(\frac{d\ln F_{i}(x)}{dx}\right)_c + \frac{3 A \alpha g (g + \gamma \frac{v_c^2}{a_c^2})a_c^4}{\gamma x_c} 
 $$
 $$- \frac{A \alpha g(g +  \gamma \frac{v_c^2}{a_c^2})a_c^4} {\gamma}\left(\frac{d\ln F_{i}(x)}{dx}\right)_c  - \frac{2 A\lambda (g + \gamma \frac{v_c^2}{a_c^2})a_c^2v_c}{x_c^2} - \left[2A\alpha g  \left(g + \gamma \frac{v_c^2}{a_c^2}\right) a_c^2+ 
\frac{(\gamma +1)v_c^2}{(\gamma -1)}\right]\left(\frac{d\Phi_{i}^{\text{eff}}}
{dx}\right)_c  = 0,
\eqno(A1)
$$ 
Using $M_c = v_c/a_c$, we get
$$
 \frac{A\alpha  \left(g+  \gamma M_c^2\right)^2a_c^4}{\gamma  x_c} + \frac{3 M_c^2 a_c^4 }{(\gamma -1)x_c} -\frac{M_c^2 a_c^4}{(\gamma -1)}\left(\frac{d\ln F_{i}(x)}{dx}\right)_c + \frac{3 A \alpha g (g + \gamma M_c^2)a_c^4}{\gamma x_c}
 $$
 $$
  - \frac{A \alpha g(g +  \gamma M_c^2)a_c^4} {\gamma}\left(\frac{d\ln F_{i}(x)}{dx}\right)_c - \frac{2 A\lambda (g + \gamma M_c^2)M_c a_c^3}{x_c^2} 
  - \left[2A\alpha g  \left(g + \gamma M_c^2 \right) a_c^2+ 
\frac{(\gamma +1)M_c^2 a_c^2}{(\gamma -1)}\right]\left(\frac{d\Phi_{i}^{\text{eff}}}
{dx}\right)_c= 0,
\eqno(A2)
$$
After some simple algebra, we have
$$
a_1 a_c^2 + a_2 a_c + a_3 = 0,
\eqno(A3)
$$
where  
\begin{align*}
a_1 =& -\frac{A\alpha (g +  \gamma M_c^2)^2}{\gamma x_c} - \frac{3 M_c^2}{(\gamma -1)x_c} + \frac{M_c^2}{(\gamma -1)}\left(\frac{d\ln F_{i}(x)}{dx}\right)_c - \frac{3A \alpha g (g +  \gamma M_c^2)}{\gamma x_c}  + \frac{A \alpha g(g + \gamma M_c^2)}
{\gamma}\left(\frac{d\ln F_{i}(x)}{dx}\right)_c, \\
a_2 =&  \frac{2A\lambda M_c (g + \gamma M_c^2)}
{x_c^2}, \\
a_3 =& \left[2A\alpha g (g + \gamma M_c^2) + \frac{(\gamma + 1)M_c^2}{(\gamma - 
	1)}\right]\left(\frac{d\Phi_{i}^{\text{eff}}}{dx}\right)_c.            
\end{align*}
In may be noted that the trivial solutions are avoided in equation (A3). Finally, we solve this equation to obtain $a_c$ and consider only positive root as $a_c>0$ always.


\end{document}